 \documentclass[twocolumn]{webofc}

\usepackage[varg]{txfonts}   
\usepackage{hyperref}
\usepackage{url}
\hypersetup{colorlinks=true,citecolor=blue,urlcolor=blue,linkcolor=blue}
\usepackage{color, soul} 

\usepackage{caption} 
\usepackage{subcaption}

\usepackage{amssymb}

\usepackage{afterpage}
\begin{document}
\title{R\&D on a high-performance electromagnetic calorimeter based on oriented crystalline scintillators}
%
%

%

\author{
\firstname{M.} \lastname{Soldani} \inst{1} \fnsep\thanks{\email{mattia.soldani@lnf.infn.it}} \and
\firstname{N.} \lastname{Argiolas} \inst{2,3} \and
\firstname{L.} \lastname{Bandiera} \inst{4} \and
\firstname{V.} \lastname{Baryshevsky} \inst{5} \and
\firstname{L.} \lastname{Bomben} \inst{6,7} \and
\firstname{C.} \lastname{Brizzolari} \inst{6,7} \and
\firstname{N.} \lastname{Canale} \inst{4} \and
\firstname{S.} \lastname{Carsi} \inst{6,7} \and
\firstname{S.} \lastname{Cutini} \inst{8} \and
\firstname{F.} \lastname{Davì }\inst{4,9} \and
\firstname{D.} \lastname{De Salvador} \inst{2,3} \and
\firstname{A.} \lastname{Gianoli} \inst{4} \and
\firstname{V.} \lastname{Guidi} \inst{10} \and
\firstname{V.} \lastname{Haurylavets} \inst{5} \and
\firstname{M.} \lastname{Korjik} \inst{5} \and
\firstname{G.} \lastname{Lezzani} \inst{6,7} \and
\firstname{A.} \lastname{Lobko} \inst{5} \and
\firstname{F.} \lastname{Longo} \inst{11,12} \and
\firstname{L.} \lastname{Malagutti} \inst{4} \and
\firstname{S.} \lastname{Mangiacavalli} \inst{6,7} \and
\firstname{V.} \lastname{Mascagna} \inst{6,7} \and
\firstname{A.} \lastname{Mazzolari} \inst{4,10} \and
\firstname{L.} \lastname{Montalto} \inst{9} \and
\firstname{P.} \lastname{Monti-Guarnieri} \inst{11,12} \and
\firstname{M.} \lastname{Moulson} \inst{1} \and
\firstname{R.} \lastname{Negrello} \inst{4,10} \and
\firstname{G.} \lastname{Paternò }\inst{4} \and
\firstname{L.} \lastname{Perna} \inst{6,7} \and
\firstname{C.} \lastname{Petroselli} \inst{6,7} \and
\firstname{M.} \lastname{Prest} \inst{6,7} \and
\firstname{D.} \lastname{Rinaldi} \inst{1,9} \and
\firstname{M.} \lastname{Romagnoni} \inst{4,10} \and
\firstname{F.} \lastname{Ronchetti} \inst{6,7} \and
\firstname{G.} \lastname{Saibene} \inst{6,7} \and
\firstname{A.} \lastname{Selmi} \inst{6,7} \and
\firstname{F.} \lastname{Sgarbossa} \inst{2,3} \and
\firstname{A.} \lastname{Sytov} \inst{4} \and
\firstname{V.} \lastname{Tikhomirov} \inst{5} \and
\firstname{E.} \lastname{Vallazza} \inst{7}
}

\institute{
INFN Laboratori Nazionali di Frascati, Frascati, Italy
\and
Università degli Studi di Padova, Padua, Italy
\and
INFN Laboratori Nazionali di Legnaro, Legnaro, Italy
\and
INFN Sezione di Ferrara, Ferrara, Italy
\and
Institute for Nuclear Problems, Belarusian State University, Minsk, Belarus
\and
Università degli Studi dell'Insubria, Como, Italy
\and
INFN Sezione di Milano Bicocca, Milan, Italy
\and
INFN Sezione di Perugia, Perugia, Italy
\and
Università Politecnica delle Marche, Ancona, Italy
\and
Università degli Studi di Ferrara, Ferrara, Italy
\and
Università degli Studi di Trieste, Trieste, Italy
\and
INFN Sezione di Trieste, Trieste, Italy
}

\abstract{%
Although inorganic scintillators are widely used in the design of electromagnetic calorimeters for high-energy physics and astrophysics, their crystalline nature and, hence, their lattice orientation are generally neglected in the detector design. However, in general, the features of the electromagnetic field experienced by the particles impinging on a crystal at a small angle with respect to a lattice axis affect their interaction mechanisms. In particular, in case of electrons/photons of $\mathcal{O} (10~\mathrm{GeV})$ or higher impinging on a high-$Z$ crystal at an angle of $\lesssim 1~\mathrm{mrad}$, the so-called strong field regime is attained: the bremsstrahlung and pair production cross sections are enhanced with respect to the case of amorphous or randomly oriented materials. Overall, the increase of these processes leads to an acceleration of the electromagnetic shower development. These effects are thoroughly investigated by the OREO (ORiEnted calOrimeter) team, and pave the way to the development of innovative calorimeters with a higher energy resolution, a higher efficiency in photon detection and an improved particle identification capabilities due to the relative boost of the electromagnetic interactions with respect to the hadronic ones. Moreover, a detector with the same resolution as the current state of the art and reduced thickness could be developed. An overview of the lattice effects at the foundation of the shower boost and of the current status of the development of an operational calorimeter prototype are presented. This concept could prove pivotal for both accelerator fixed-target experiments and satellite-borne 
$\gamma$-ray observatories.
}
\maketitle

\section{Introduction}
\label{intro}

Crystalline scintillators are extensively used in the design of electromagnetic calorimeters (ECALs) for existing and future experiments in high-energy physics and astroparticle physics. Indeed, homogeneous calorimeters based on inorganic crystals feature full shower containment in a compact detector volume, owing to the high $Z$ and density, unparalleled energy resolution, generally fast response, rather high resistance to radiation-induced damage. Moreover, scintillating crystals are in general easy to machine, allowing for blocks with size $\mathcal{O} (\mathrm{cm})$ to be obtained. This allows for fine transverse segmentation and, as recently discussed, e.g., in \cite{CRILIN}, longitudinal sampling to be achieved.

The development of an electromagnetic shower in a crystal calorimeter is conventionally modelled as occurring in an amorphous medium, in which only mutually independent interactions with the Coulomb field of single atoms are considered. This picture neglects the significant role that the properly oriented crystalline lattice has on the electromagnetic interactions. Indeed, when a particle moves close to a periodic array of atoms (namely, an axis) in the lattice, it experiences an electromagnetic field that is equivalent to the coherent sum of many contributions from all the single atoms and is approximately constant along the string direction. As a result, the electromagnetic processes the high-energy particle impinging on the crystal undergo can be significantly altered \cite{baier1998electromagnetic,uggerhoj2005}. In the following, the features of the coherent effects occurring upon interaction of high-energy (i.e., at the GeV scale and higher) electrons/positrons and photons in lead tungstate (PWO) are outlined. The implications of the existence of these effects in calorimetry are also discussed.

Some of the experimental results most recently obtained probing various PWO samples (aligned to two of the strongest axes, i.e., $\langle001\rangle$ and $\langle100\rangle$ -- along which PWO crystals are generally grown) with different-energy electron beams at the CERN H2 ($100$-$120~\mathrm{GeV}$) and DESY T21 ($5.6~\mathrm{GeV}$) beamlines are shown. The measurement setup is described in detail, e.g., in \cite{soldani2022_pwo,selmi2023,soldanithesis}. Essentially, single electrons impinging on the axially aligned crystalline sample (installed on a high-precision goniometer) were tracked; a bending magnet separated the electromagnetic radiation from the charged particles emerging from the sample, and the total energy of the former was measured with an electromagnetic calorimeter (“$\gamma$-CAL”).

\section{High-energy coherent effects in oriented PWO}
\label{sec1}

At sufficiently high initial energy $E_0$, Lorentz contraction occurs and the lattice field in the particle rest frame is boosted up to a strength larger than the Schwinger critical field ($\mathcal{E}_0 \sim 1.32 \cdot 10^{16}~\mathrm{V/cm}$). The so-called Strong Field (SF) regime is attained \cite{baier1998electromagnetic,uggerhoj2005}: the effect of such an intense field on the incident electron/positron is the enhancement of the radiation emission probability with respect to the Bethe-Heitler description typical of amorphous media \cite{kimball1985}. By crossing symmetry, in case of an incident high-energy photon, the pair production probability is greatly increased as well \cite{Baryshevskii:1983JETP,BARYSHEVSKII1985335}. Overall, the increase of these processes leads to an acceleration of the first stages of the electromagnetic shower development with respect to the case of an amorphous or randomly oriented medium \cite{bandiera2018,2024_Soldani_2_ARXIV}.

\begin{figure}[t]
\centering
\includegraphics[width=\columnwidth,clip]{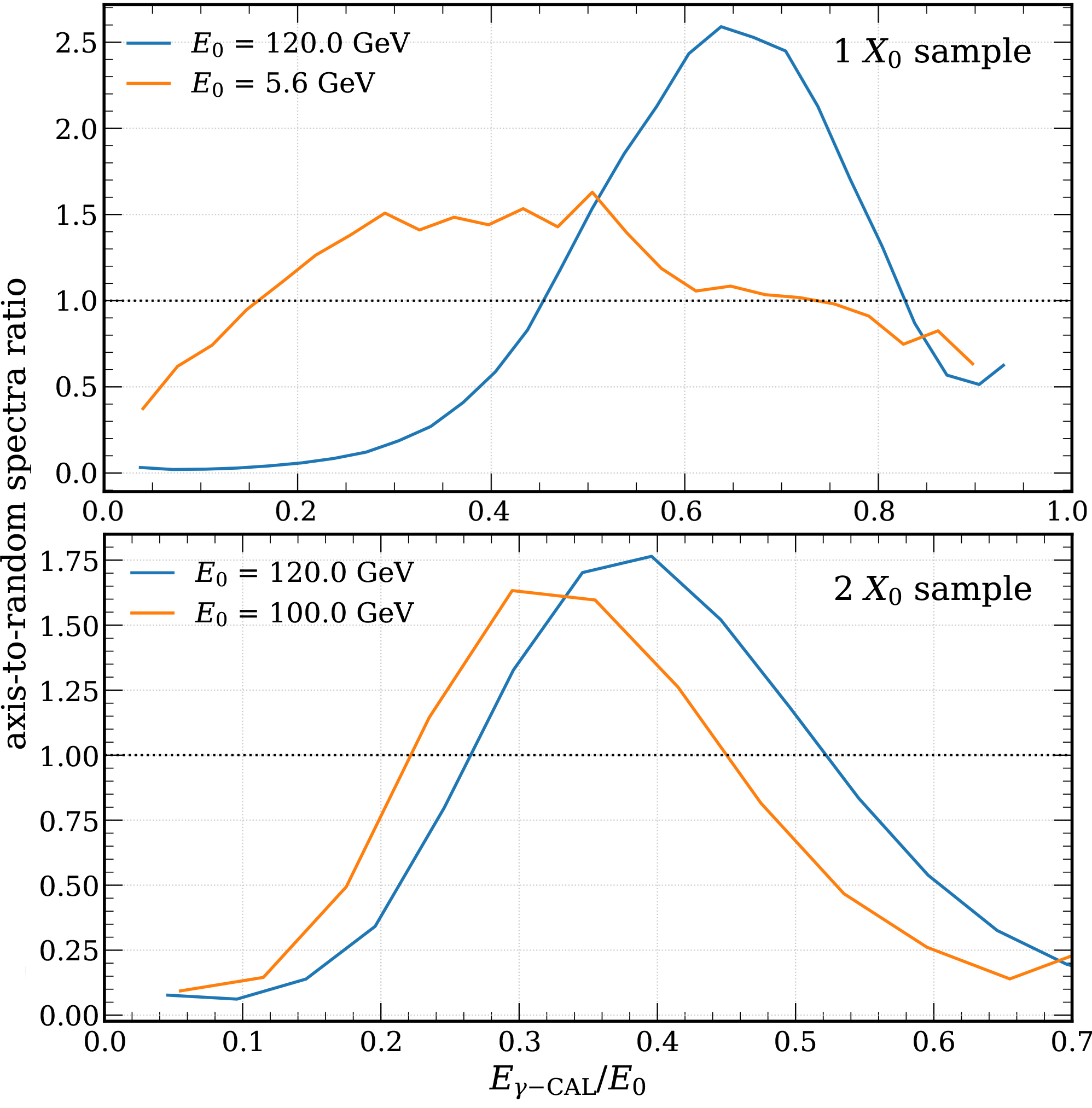}
\caption{Axis-to-random ratio of the spectra of the energy (normalised to the primary energy) of the electromagnetic radiation emerging from a $1~X_0$ \textit{(top)} and a $2~X_0$ \textit{(bottom)} PWO sample (and measured with the $\gamma$-CAL) probed with electrons at different primary energies.}
\label{fig:energies}
\end{figure}

The condition for the SF regime to be attained can be written in terms of the dimensionless parameter $\chi = \gamma \mathcal{E}_\mathrm{lab} / \mathcal{E}_0$, where $\gamma = E_0 / m_e c^2$ is the Lorentz factor and $\mathcal{E}_\mathrm{lab}$ is the axial field in the lab frame \cite{baier1998electromagnetic,uggerhoj2005}. The energy threshold for the onset of the full SF regime can then be determined by setting $\chi \geq 1$. For the $\langle001\rangle$ and $\langle100\rangle$ axes of PWO, this corresponds to an initial energy $E_0 \gtrsim 25~\mathrm{GeV}$ \cite{2024_Soldani_2_ARXIV}. Limited SF effects in crystals are observed down to a primary energy of about an order of magnitude less \cite{2021_Soldani,soldani2024}, i.e., a few $\mathrm{GeV}$ in case of PWO.

Figure \ref{fig:energies} shows the ratios between the radiation energy spectra measured with the $\gamma$-CAL under axial and amorphous-like (i.e., at very large angle -- several degrees -- with respect to the axis) alignment, for two PWO samples, $1~X_0$ and $2~X_0$ thick respectively, and for different energies in both full and limited SF regime.

Some conclusions can be drawn: firstly, for higher $E_0$, the axis-to-random radiation enhancement peak is higher and located at a higher fraction of $E_0$, i.e., a harder component of the radiation energy spectrum is boosted; this reflects the facts that the SF effect strength grows with $E_0$ and that a higher $\chi$ results in the emission of harder synchrotron-like radiation \cite{bandiera2018,soldani2023}.

Moreover, the maximum enhancement and the fraction of $E_0$ at which it occurs decrease in thicker samples. Indeed, in case of multi-$X_0$ crystals, the effect of the SF is reduced as the electromagnetic shower initiated by the primary particle develops and more lower-energy, off-axis secondaries are produced. Asymptotically, this makes the total output radiation state (which results from the interactions occurring throughout the whole sample thickness) from a crystal of several $X_0$ become more similar to that obtained from an amorphous target.

Another important parameter at play in determining the strength of the SF effects on the interactions of an incident particle is the so-called misalignment angle $\theta_{\text{mis}}$, i.e., the angle between the particle trajectory and the lattice axis. An estimate of angle required for the shower development acceleration to be attained is provided by $\Theta_0 = U_0 / m_e c^2$ \cite{uggerhoj2005}, $U_0$ being the axis potential. In case of both the $\langle001\rangle$ and $\langle100\rangle$ axes of PWO, $\Theta_0 \sim 0.9~\mathrm{mrad}$ \cite{2024_Soldani_2_ARXIV}. At a primary energy of few $\mathrm{GeV}$ or higher, this value is significantly larger than the (energy-dependent) angular acceptance of the well known channelling of charged particles between neighbouring planes/axes \cite{lindhard65}, i.e. $\psi_\mathrm{L} = \sqrt{ 2 U_0 / E_0 }$, which is of the order of $0.4~\mathrm{mrad}$ ($0.08~\mathrm{mrad}$) for $E_0 = 5.6~\mathrm{GeV}$ ($120~\mathrm{GeV}$) \cite{soldanithesis}.

For $\theta_{\text{mis}} > \Theta_0$, the enhancement of the electromagnetic processes gradually decreases and other lattice-borne effects occur, such as coherent bremsstrahlung and coherent pair production \cite{dyson_anisotropy_1955, diambrini-palazzi_interazioni_1962}. It has been observed that such phenomena determine a lesser increase of both bremsstrahlung and pair production for $\theta_\mathrm{mis}$ up to about $1^\circ$ \cite{2024_Soldani_2_ARXIV}.

\begin{figure}[t]
\centering
\includegraphics[width=\columnwidth, clip]{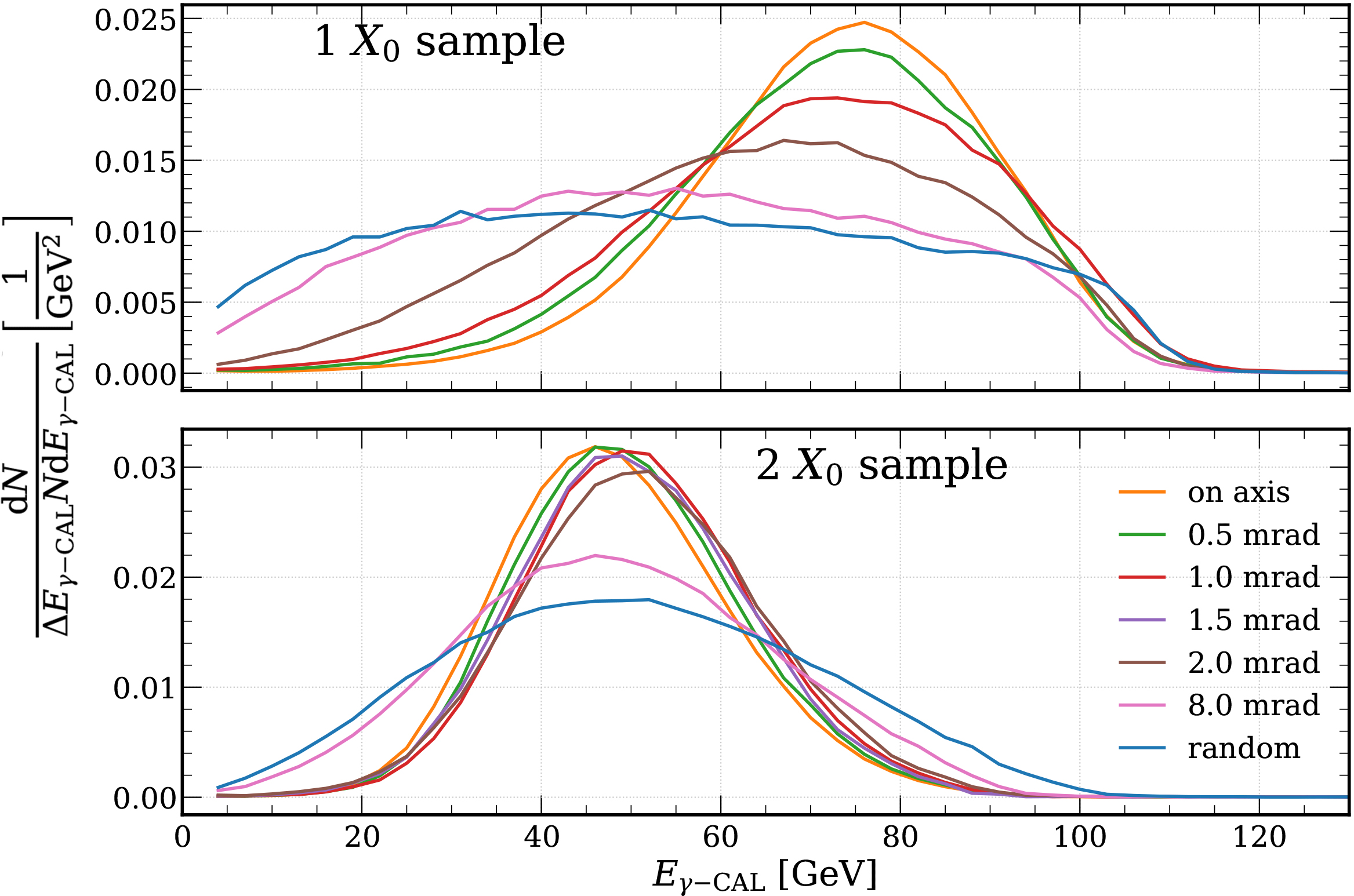}
\caption{Distributions of the energy of the electromagnetic radiation emerging from a $1~X_0$ \textit{(top)} and a $2~X_0$ \textit{(bottom)} PWO sample (and measured with the $\gamma$-CAL) probed with $120$-$\mathrm{GeV}/c$ electrons at different misalignment angles.}
\label{fig:transition}
\end{figure}

The features of the full-SF ($E_0 = 120~\mathrm{GeV}$) radiation energy spectrum throughout a transition between on-axis and amorphous-like configurations are clearly shown in figure \ref{fig:transition} for the same $1~X_0$ and $2~X_0$ samples mentioned above (in figure \ref{fig:energies}). Considering the $1~X_0$ sample, the features of the spectrum measured at $\theta_{\text{mis}} = 2~\mathrm{mrad} \sim 2 \Theta_0$ are in between those of the spectra measured in axial and in random orientation. On the other hand, the spectra measured with the $2~X_0$ crystal show approximately the same features from the axis up to $2~\mathrm{mrad}$, hinting at a larger angular acceptance with respect to the thinner sample.  This might partially reflect the fact that the $2~X_0$ sample, for which a higher surface mosaicity than the $1~X_0$ one was measured \cite{soldanithesis}, has a higher internal mosaicity as well. Moreover, as discussed above, a higher number of lower-energy secondary particles is generated in thicker samples, which, recalling that the range of channelling, $\psi_\mathrm{L}$ (also proportional to that of coherent bremsstrahlung \cite{soldanithesis}), grows as the particle energy decreases, might contribute to the observed direct relation between the acceptance of the coherent interactions and the sample thickness as well.

\section{Impact in calorimetry}
\label{sec2}

As shown in section \ref{sec1}, the SF-related effects to the electromagnetic interactions are macroscopic, therefore it is important that they are considered in the design of a crystal calorimeter. On one hand, the shower development acceleration may prove beneficial in developing next-generation ECALs that operate in forward-geometry configuration at the multi-$\mathrm{GeV}$ scale and above. Indeed, these effects lead to better shower containment in the calorimeter volume, which would result in an improvement of the energy resolution. Alternatively, it would be possible to build an ECAL with the same resolution as currently achievable but with reduced thickness. Moreover, the hadronic interactions are not affected by the crystalline lattice, which could be exploited to develop a highly compact oriented crystal calorimeter with enhanced transparency to the passage of hadrons (see, e.g., \cite{hike_loi, hike}) and, if segmented, with more $\gamma$/hadron discrimination power. These concepts could prove particularly appealing in case of a $\gamma$-ray detector installed on a satellite with pointing capabilities and of an ECAL and/or a preshower detector in a fixed-target experiment or in the forward region of a collider experiment. Further details can be found in \cite{Bandiera23,2024_Soldani_2_ARXIV} and references therein.

On the other hand, particular care should be taken to the lattice orientation of calorimeter towers in general, as (some of) the particles impinging on the detector might be aligned with a crystalline axis by chance. In fact, it was observed that PWO crystals, such as those of the CMS ECAL \cite{17_lecoq}, are grown at a small angle with respect to one of the main axes. Preliminary measurements performed on a spare block of the CMS endcap indicate a miscut (i.e., the misalignment between the direction orthogonal to one of the block faces and the nearest lattice axis) $\lesssim 0.3^\circ$ and a mosaicity of $\sim 200~\mu\mathrm{rad}$. 

\begin{figure*}[t]
\centering
\includegraphics[width=1.1\columnwidth]{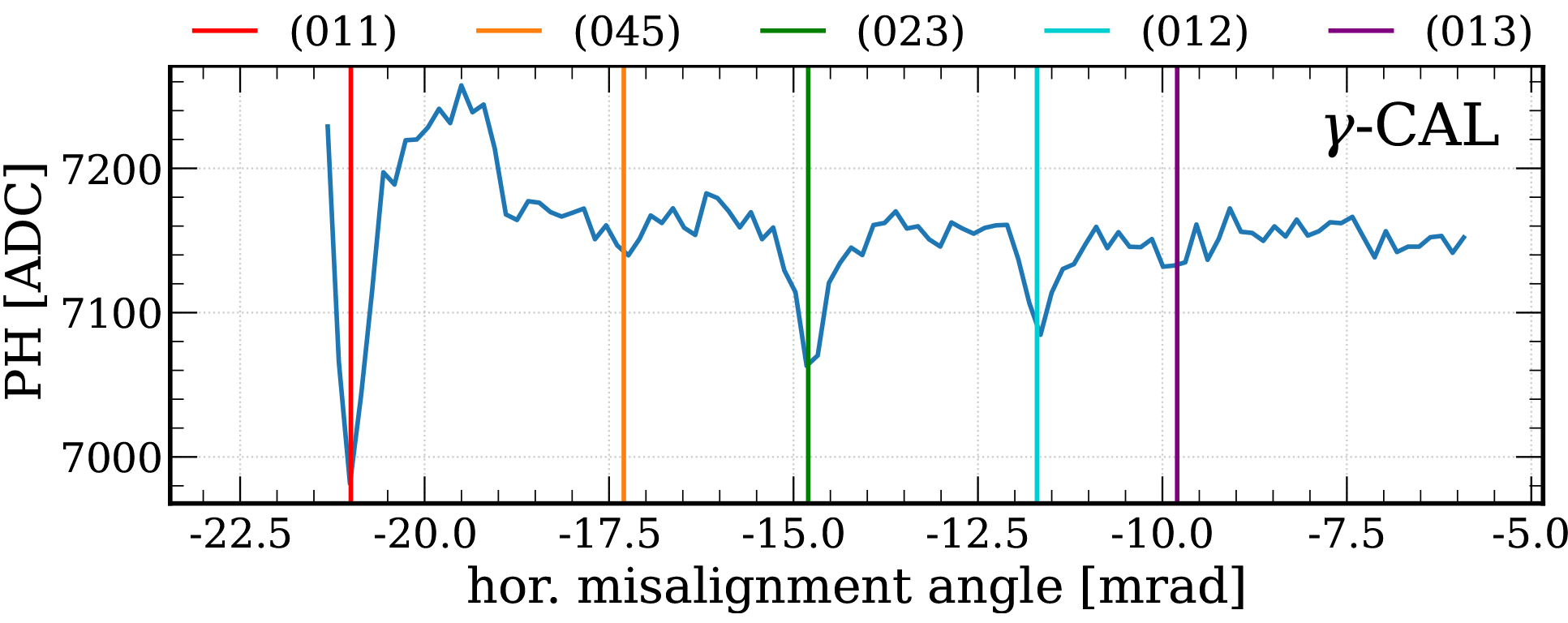}
\caption{Response of the $\gamma$-CAL (here detecting both the electromagnetic radiation and the output charged particles) to $120$-$\mathrm{GeV}/c$ electrons measured in a goniometer scan over an angular region with several high-order planes of a $4.6~X_0$ PWO sample.}
\label{fig:sample_scan}
\end{figure*}

The influence of the coherent effects is expected to be limited in case of a calorimeter with a thickness of $\gtrsim 20~X_0$ and without longitudinal segmentation, since the primary energy is fully deposited in the active volume regardless of the lattice orientation. Conversely, the performance of the upstream layer(s) of a calorimeter with longitudinally segmented towers could be significantly affected by the shower boost discussed in this work if the incident particles impinge on different points of the detector front and with different misalignment angles. Moreover, as recent measurements performed with $120$-$\mathrm{GeV}/c$ electrons on a crystal with a thickness of $4.6~X_0$ show, macroscopic modifications to the amorphous-like behaviour occur not only with the axes with the strongest fields, but also in presence of higher-order axes and planes -- i.e., distributions of atoms arranged in a periodic, 2-dimensional array. An example is provided in figure \ref{fig:sample_scan}, which demonstrates that even rather low-potential planes induce a non-negligible modification to the features of the crystal output state observed by the $\gamma$-CAL -- which in this measurement includes both photons and charged particles, as the bending magnet was not active.

\section{Conclusions}
\label{concl}

The modifications to the electromagnetic interactions occurring in axially oriented crystals at high energy are macroscopic, and could be exploited to significantly improve the performance of homogeneous calorimeters. This is the primary objective of the OREO (ORientEd calOrimeter) project, which aims at the development of PWO calorimeter prototype with longitudinal segmentation and an axially aligned upstream layer by 2025 \cite{Bandiera23}. Extensive studies should be performed on both existing and future crystal calorimeters to improve the level of control over the lattice orientation.

\section*{Acknowledgements}

This work was primarily funded by INFN CSN5 through the STORM project. We also acknowledge partial support of INFN CSN5 (OREO and Geant4-INFN projects) and CSN1 (NA62 experiment; RD-FLAVOUR project), of the Italian Ministry of University and Research (PRIN 2022Y87K7X) and of the European Commission (Horizon 2020 MSCA-RISE N-LIGHT, GA 872196; Horizon 2020 AIDAinnova, GA 101004761; Horizon 2020 MSCA IF Global TRILLION, GA 101032975; Horizon EIC Pathfinder Open TECHNO-CLS, GA 101046458). 

\bibliography{bibliography}

\begin{thebibliography}{21}

\bibitem{CRILIN}
S.~Ceravolo et~al., Crilin: A crystal calorimeter with longitudinal information for a future muon collider, J. Instrum. \textbf{17}, P09033 (2022). \doiwoc{10.1088/1748-0221/17/09/P09033}

\bibitem{baier1998electromagnetic}
V.N. Baier et~al., Electromagnetic processes at high energies in oriented single crystals (World Scientific, 1998)

\bibitem{uggerhoj2005}
U.I. Uggerh\o{}j, {The interaction of relativistic particles with strong crystalline fields}, Rev. Mod. Phys. \textbf{77}, 1131 (2005). \doiwoc{10.1103/RevModPhys.77.1131}

\bibitem{soldani2022_pwo}
M.~Soldani et~al., A high-performance custom photodetection system to probe the light yield enhancement in oriented crystals, J. Phys. Conf. Ser. \textbf{2374}, 012112 (2022). \doiwoc{10.1088/1742-6596/2374/1/012112}

\bibitem{selmi2023}
A.~Selmi et~al., {Experimental layout for the direct measurement of electromagnetic shower acceleration in an oriented crystal scintillator}, Nucl. Instrum. Methods Phys. Res. A \textbf{1048}, 167948 (2023). \doiwoc{10.1016/j.nima.2022.167948}

\bibitem{soldanithesis}
M.~Soldani, Ph.D. thesis, Università degli Studi di Ferrara (2023), \urlstyle{tt}\url{https://cds.cern.ch/record/2864634}

\bibitem{kimball1985}
J.C. Kimball et~al., Quantum electrodynamics and channeling in crystals, Phys. Rep. \textbf{125}, 69 (1985). \doiwoc{10.1016/0370-1573(85)90021-3}

\bibitem{Baryshevskii:1983JETP}
V.G. Baryshevskii et~al., Creation of transversely polarized high-energy electrons and positrons in crystals, Sov. Phys. JETP \textbf{58}, 135 (1983).

\bibitem{BARYSHEVSKII1985335}
V.G. Baryshevskii et~al., Pair production in a slowly varying electromagnetic field and the pair production process, Phys. Lett. A \textbf{113}, 335 (1985). \doiwoc{https://doi.org/10.1016/0375-9601(85)90178-1}

\bibitem{bandiera2018}
L.~Bandiera et~al., Strong reduction of the effective radiation length in an axially oriented scintillator crystal, Phys. Rev. Lett. \textbf{121}, 021603 (2018). \doiwoc{10.1103/PhysRevLett.121.021603}

\bibitem{2024_Soldani_2_ARXIV}
M.~Soldani et~al., {Acceleration of electromagnetic shower development and enhancement of light yield in oriented scintillating crystals} (2024), \texttt{2404.12016}.

\bibitem{2021_Soldani}
M.~Soldani et~al., {Next-generation ultra-compact calorimeters based on oriented crystals}, PoS \textbf{ICHEP2020}, 872 (2021). \doiwoc{10.22323/1.390.0872}

\bibitem{soldani2024}
M.~Soldani et~al., Radiation in oriented crystals: Innovative application to future positron sources, Nucl. Instrum. Methods Phys. Res. A \textbf{1058}, 168828 (2024). \doiwoc{https://doi.org/10.1016/j.nima.2023.168828}

\bibitem{soldani2023}
M.~Soldani et~al., {Strong enhancement of electromagnetic shower development induced by high-energy photons in a thick oriented tungsten crystal}, Eur. Phys. J. C \textbf{83}, 101 (2023). \doiwoc{10.1140/epjc/s10052-023-11247-x}

\bibitem{lindhard65}
J.~Lindhard, {Influence of crystal lattice on motion of energetic charged particles}, Kongel. Dan. Vidensk. Selsk. Mat. Fys. Medd. \textbf{34}, 1 (1965).

\bibitem{dyson_anisotropy_1955}
F.J. Dyson, H.~\"Uberall, Anisotropy of bremsstrahlung and pair production in single crystals, Phys. Rev. \textbf{99}, 604 (1955). \doiwoc{10.1103/PhysRev.99.604}

\bibitem{diambrini-palazzi_interazioni_1962}
G.~Diambrini-Palazzi, Interazioni di fotoni ed elettroni di alta energia in cristalli, Nuovo Cim. \textbf{25}, 88 (1962). \doiwoc{10.1007/BF02860173}

\bibitem{hike_loi}
{HIKE Collaboration}, {HIKE, High Intensity Kaon Experiments at the CERN SPS: Letter of Intent} (2022), \urlstyle{tt}\url{https://cds.cern.ch/record/2839661}

\bibitem{hike}
{HIKE Collaboration}, {High Intensity Kaon Experiments (HIKE) at the CERN SPS: Proposal for Phases 1 and 2} (2023), \urlstyle{tt}\url{https://cds.cern.ch/record/2878543}

\bibitem{Bandiera23}
L.~Bandiera et~al., A highly-compact and ultra-fast homogeneous electromagnetic calorimeter based on oriented lead tungstate crystals, Front. Phys. \textbf{11}, 10.3389/fphy.2023.1254020 (2023). \doiwoc{10.3389/fphy.2023.1254020}

\bibitem{17_lecoq}
P.~Lecoq, A.~Getkin, M.~Korzhik, {Inorganic Scintillators for Detector Systems} (Springer, 2017)

\end{thebibliography}

\end{document}